\documentclass[aps,prl,twocolumn,superscriptaddress,showpacs,preprintnumbers,amsmath,amssymb, linenumbers]{revtex4}
\usepackage{graphicx,color} 
\usepackage{dcolumn} 
\usepackage{subfigure}
\usepackage{color,soul}
\usepackage{multirow}
\usepackage{braket}
\RequirePackage{xspace}
\def\Dbar{\kern 0.2em\overline{\kern -0.2em D}{}\xspace}
\def\Kbar{\kern 0.2em\overline{\kern -0.2em K}{}\xspace}
\def\Bbar{\kern 0.2em\overline{\kern -0.2em B}{}\xspace}
\def\Dzb {\ensuremath{\Dbar^0}\xspace}
\RequirePackage{xspace}
\graphicspath{{ps}}

\begin{document}

\title{ \quad\\[1.0cm] {\boldmath Measurement of the charm-mixing parameter $y_{CP}$ in $D^{0}\to K^{0}_{S}\omega$ decays at Belle} }

\noaffiliation
\affiliation{University of the Basque Country UPV/EHU, 48080 Bilbao}
\affiliation{Beihang University, Beijing 100191}
\affiliation{Brookhaven National Laboratory, Upton, New York 11973}
\affiliation{Budker Institute of Nuclear Physics SB RAS, Novosibirsk 630090}
\affiliation{Faculty of Mathematics and Physics, Charles University, 121 16 Prague}
\affiliation{Chonnam National University, Gwangju 61186}
\affiliation{University of Cincinnati, Cincinnati, Ohio 45221}
\affiliation{Deutsches Elektronen--Synchrotron, 22607 Hamburg}
\affiliation{Duke University, Durham, North Carolina 27708}
\affiliation{Key Laboratory of Nuclear Physics and Ion-beam Application (MOE) and Institute of Modern Physics, Fudan University, Shanghai 200443}
\affiliation{II. Physikalisches Institut, Georg-August-Universit\"at G\"ottingen, 37073 G\"ottingen}
\affiliation{SOKENDAI (The Graduate University for Advanced Studies), Hayama 240-0193}
\affiliation{Gyeongsang National University, Jinju 52828}
\affiliation{Department of Physics and Institute of Natural Sciences, Hanyang University, Seoul 04763}
\affiliation{University of Hawaii, Honolulu, Hawaii 96822}
\affiliation{High Energy Accelerator Research Organization (KEK), Tsukuba 305-0801}
\affiliation{J-PARC Branch, KEK Theory Center, High Energy Accelerator Research Organization (KEK), Tsukuba 305-0801}
\affiliation{Forschungszentrum J\"{u}lich, 52425 J\"{u}lich}
\affiliation{IKERBASQUE, Basque Foundation for Science, 48013 Bilbao}
\affiliation{Indian Institute of Science Education and Research Mohali, SAS Nagar, 140306}
\affiliation{Indian Institute of Technology Bhubaneswar, Satya Nagar 751007}
\affiliation{Indian Institute of Technology Guwahati, Assam 781039}
\affiliation{Indian Institute of Technology Hyderabad, Telangana 502285}
\affiliation{Indian Institute of Technology Madras, Chennai 600036}
\affiliation{Indiana University, Bloomington, Indiana 47408}
\affiliation{Institute of High Energy Physics, Chinese Academy of Sciences, Beijing 100049}
\affiliation{Institute of High Energy Physics, Vienna 1050}
\affiliation{Institute for High Energy Physics, Protvino 142281}
\affiliation{INFN - Sezione di Napoli, 80126 Napoli}
\affiliation{INFN - Sezione di Torino, 10125 Torino}
\affiliation{Advanced Science Research Center, Japan Atomic Energy Agency, Naka 319-1195}
\affiliation{J. Stefan Institute, 1000 Ljubljana}
\affiliation{Institut f\"ur Experimentelle Teilchenphysik, Karlsruher Institut f\"ur Technologie, 76131 Karlsruhe}
\affiliation{Department of Physics, Faculty of Science, King Abdulaziz University, Jeddah 21589}
\affiliation{Kitasato University, Sagamihara 252-0373}
\affiliation{Korea Institute of Science and Technology Information, Daejeon 34141}
\affiliation{Korea University, Seoul 02841}
\affiliation{Kyoto University, Kyoto 606-8502}
\affiliation{Kyungpook National University, Daegu 41566}
\affiliation{LAL, Univ. Paris-Sud, CNRS/IN2P3, Universit\'{e} Paris-Saclay, Orsay 91898}
\affiliation{\'Ecole Polytechnique F\'ed\'erale de Lausanne (EPFL), Lausanne 1015}
\affiliation{P.N. Lebedev Physical Institute of the Russian Academy of Sciences, Moscow 119991}
\affiliation{Faculty of Mathematics and Physics, University of Ljubljana, 1000 Ljubljana}
\affiliation{Ludwig Maximilians University, 80539 Munich}
\affiliation{University of Maribor, 2000 Maribor}
\affiliation{Max-Planck-Institut f\"ur Physik, 80805 M\"unchen}
\affiliation{School of Physics, University of Melbourne, Victoria 3010}
\affiliation{University of Mississippi, University, Mississippi 38677}
\affiliation{University of Miyazaki, Miyazaki 889-2192}
\affiliation{Moscow Physical Engineering Institute, Moscow 115409}
\affiliation{Moscow Institute of Physics and Technology, Moscow Region 141700}
\affiliation{Graduate School of Science, Nagoya University, Nagoya 464-8602}
\affiliation{Universit\`{a} di Napoli Federico II, 80055 Napoli}
\affiliation{Nara Women's University, Nara 630-8506}
\affiliation{National Central University, Chung-li 32054}
\affiliation{National United University, Miao Li 36003}
\affiliation{Department of Physics, National Taiwan University, Taipei 10617}
\affiliation{H. Niewodniczanski Institute of Nuclear Physics, Krakow 31-342}
\affiliation{Nippon Dental University, Niigata 951-8580}
\affiliation{Niigata University, Niigata 950-2181}
\affiliation{University of Nova Gorica, 5000 Nova Gorica}
\affiliation{Novosibirsk State University, Novosibirsk 630090}
\affiliation{Osaka City University, Osaka 558-8585}
\affiliation{Pacific Northwest National Laboratory, Richland, Washington 99352}
\affiliation{Panjab University, Chandigarh 160014}
\affiliation{Peking University, Beijing 100871}
\affiliation{University of Pittsburgh, Pittsburgh, Pennsylvania 15260}
\affiliation{Punjab Agricultural University, Ludhiana 141004}
\affiliation{Theoretical Research Division, Nishina Center, RIKEN, Saitama 351-0198}
\affiliation{University of Science and Technology of China, Hefei 230026}
\affiliation{Seoul National University, Seoul 08826}
\affiliation{Showa Pharmaceutical University, Tokyo 194-8543}
\affiliation{Soochow University, Suzhou 215006}
\affiliation{Soongsil University, Seoul 06978}
\affiliation{University of South Carolina, Columbia, South Carolina 29208}
\affiliation{Sungkyunkwan University, Suwon 16419}
\affiliation{School of Physics, University of Sydney, New South Wales 2006}
\affiliation{Department of Physics, Faculty of Science, University of Tabuk, Tabuk 71451}
\affiliation{Tata Institute of Fundamental Research, Mumbai 400005}
\affiliation{Department of Physics, Technische Universit\"at M\"unchen, 85748 Garching}
\affiliation{Toho University, Funabashi 274-8510}
\affiliation{Department of Physics, Tohoku University, Sendai 980-8578}
\affiliation{Earthquake Research Institute, University of Tokyo, Tokyo 113-0032}
\affiliation{Department of Physics, University of Tokyo, Tokyo 113-0033}
\affiliation{Tokyo Institute of Technology, Tokyo 152-8550}
\affiliation{Tokyo Metropolitan University, Tokyo 192-0397}
\affiliation{Virginia Polytechnic Institute and State University, Blacksburg, Virginia 24061}
\affiliation{Wayne State University, Detroit, Michigan 48202}
\affiliation{Yamagata University, Yamagata 990-8560}
\affiliation{Yonsei University, Seoul 03722}
  \author{M.~Nayak}\affiliation{School of Physics and Astronomy, Tel Aviv University, Tel Aviv 69978} 
  \author{D.~Cinabro}\affiliation{Wayne State University, Detroit, Michigan 48202} 
  \author{I.~Adachi}\affiliation{High Energy Accelerator Research Organization (KEK), Tsukuba 305-0801}\affiliation{SOKENDAI (The Graduate University for Advanced Studies), Hayama 240-0193} 
  \author{H.~Aihara}\affiliation{Department of Physics, University of Tokyo, Tokyo 113-0033} 
  \author{S.~Al~Said}\affiliation{Department of Physics, Faculty of Science, University of Tabuk, Tabuk 71451}\affiliation{Department of Physics, Faculty of Science, King Abdulaziz University, Jeddah 21589} 
  \author{D.~M.~Asner}\affiliation{Brookhaven National Laboratory, Upton, New York 11973} 
  \author{H.~Atmacan}\affiliation{University of South Carolina, Columbia, South Carolina 29208} 
  \author{T.~Aushev}\affiliation{Moscow Institute of Physics and Technology, Moscow Region 141700} 
  \author{R.~Ayad}\affiliation{Department of Physics, Faculty of Science, University of Tabuk, Tabuk 71451} 
  \author{V.~Babu}\affiliation{Deutsches Elektronen--Synchrotron, 22607 Hamburg} 
% ...
\collaboration{The Belle Collaboration}

\begin{abstract}
We report the first measurement of the charm-mixing parameter $y_{CP}$ in $D^{0}$ decays to the $CP$-odd final state $K^{0}_{S}\omega$. The study uses the full Belle $e^{+}e^-$ annihilation data sample of $976\rm~ fb^{-1}$ taken at or near the $\Upsilon(4S)$ centre-of-mass energy. 
We find $y_{CP} = (0.96 \pm 0.91 \pm 0.62 {}^{+0.17}_{-0.00})\%$, where the first uncertainty is statistical, the second is systematic due to event selection and background, and the last is due to possible presence of $CP$-even decays in the data sample.
\end{abstract}
\pacs{12.15.Ff, 13.25.Ft, 14.40.Lb}
\maketitle
\tighten
{\renewcommand{\thefootnote}{\fnsymbol{footnote}}}
\setcounter{footnote}{0}

In systems of neutral mesons and antimesons, flavor-changing weak interactions induce mixing. 
The mixing phenomenon originates due to the difference between mass and flavor eigenstates and has been observed in the $K^{0}-\Kbar^{0}$, $B_{(d,s)}^{0}-\Bbar_{(d,s)}^{0}$, and $D^{0}-\Dzb$ systems~\cite{pdg}. 
In the latter case, the mass eigenstates $\ket{D_{1,2}}$ with masses $m_{1,2}$ and widths $\Gamma_{1,2}$ can be expressed as linear combinations of the flavor eigenstates,
\begin{equation}
  \ket{D_{1,2}} =  p\ket{D^{0}} \pm q\ket{\Dzb}, 
\end{equation}
with $|p|^{2} + |q|^{2} =1$. The mixing rate is characterized by two parameters: $x = \Delta m/\Gamma$ and $y = \Delta \Gamma/{2\Gamma}$. 
Here $\Delta m = m_{2} - m_{1}$ and $\Delta \Gamma = \Gamma_{2} - \Gamma_{1}$ are the differences 
in mass and decay width, respectively, and $\Gamma = (\Gamma_{2} + \Gamma_{1})/2$ 
is the average decay width of the two mass eigenstates. 
If $CP$ is conserved, $p = q = 1/{\sqrt 2}$, and the mass eigenstates $\ket{D_{1,2}}$ coincide with $CP$-odd $(D_-)$ and -even $(D_+)$ states, respectively. Here the phase convention is chosen such that
$CP\ket{D^{0}} = -\ket{\Dzb}$ and $CP\ket{\Dzb} = -\ket{D^0}$.

For small values of the mixing parameters, $|x|$, $|y|$ $\ll$ 1, the decay-time dependence of initially produced $D^0$ and $\Dzb$ mesons decaying to a $CP$ eigenstate is approximately exponential. The effective lifetime here differs from that in decays to flavor eigenstates such as $D^0 \to K^-\pi^+$~\cite{conjugate}. Summing $D^0$ and $\Dzb$ decays, the time-dependent decay rate to a $CP$ eigenstate can be written as
\begin{equation}
\frac{d\Gamma(D^0 \to f_\pm) + d\Gamma(\Dzb \to f_\pm)}{dt} \propto e^{-\Gamma (1 + \eta_{f}y_{CP}) t},
\end{equation}
where $\eta_{f} = +1~(-1)$ for $CP$-even (-odd) final states. Neglecting possible $CP$ violation in decays, $y_{CP}$ is related to $x$ and $y$ as
\begin{equation}
y_{CP} = \frac{1}{2}\left(\left|\frac{q}{p}\right|+\left|\frac{p}{q}\right|\right) y \cos\phi - \frac{1}{2}\left(\left|\frac{q}{p}\right|-\left|\frac{p}{q}\right|\right) x \rm \sin\phi,
\label{eq:ycp2}
\end{equation}
where $\phi = \arg(q/p)$. In the limit of $CP$ conservation $(|q/p|=1, \phi=0)$, $y_{CP} = y$. 
Note that $y_{CP}$ also depends on $CP$ violation in decay, making the difference in $y_{CP}$ between $CP$-even and -odd final states sensitive to $CP$ violation in decay~\cite{Gersabeck}.

The most precise measurement of $y_{CP}$ has been performed with decays to $CP$-even final states $K^{+}K^{-}$ and $\pi^{+}\pi^{-}$ \cite{staric_2016, lees, lhcb_ycp}. A mixing search in $CP$-odd decays was also performed by Belle using $\rm 673~fb^{-1}$ data in $D^{0}\to K^{0}_{S}K^{+}K^{-}$ \cite{zupanc} by comparing the effective lifetimes in $CP$-even and -odd components of this final state and assuming $|q/p|=1$. The current world average value of $y_{CP}$ is $(0.715 \pm 0.111)\%$ \cite{hflav}. 

In this paper, we search for $D$-mixing in the $CP$-odd decay $D^{0} \to K^{0}_{S}\omega$ with $\omega \to \pi^{+}\pi^{-}\pi^{0}$. This decay is favorable as it has a relatively large branching fraction 
of $(0.99 \pm 0.05) \%$ ~\cite{pdg}, nearly five times that of $D^{0} \to K^{0}_{S}\phi$, and the two charged tracks from the $D^0$ decay vertex allow for an accurate measurement of the $D^0$ decay time.  The narrowness of the $\omega$ peak leads to small contamination by other resonant or non-resonant decays to the $D^0 \to K^{0}_S \pi^+ \pi^- \pi^0$ final state. We extract $y_{CP}$ by comparing the lifetimes of $K^{0}_{S}\omega$ and $K^{-}\pi^{+}$.
Since $d\Gamma(D^0 \to K^{-}\pi^{+})/dt \propto e^{-\Gamma t}$, Eq. (2) implies
\begin{equation}
 y_{CP} = 1 - \frac{\Gamma(K^{0}_{S}\omega)}{\Gamma(K^{-}\pi^+)} 
 = 1 - \frac{\tau(K^{-}\pi^{+})}{\tau(K^{0}_{S}\omega)}\,.
\end{equation}

Our study is based on the full data sample of $976~\rm fb^{-1}$ recorded with the Belle~\cite{belle}
detector at the KEKB asymmetric-energy $e^{+}e^{-}$ collider~\cite{kekb} at a center-of-mass energy near the $\Upsilon(4S)$ resonance. The detector components relevant for this work are a silicon vertex detector (SVD), a 50-layer central drift chamber (CDC), and an electromagnetic calorimeter (ECL) comprised of CsI(Tl) crystals; all located inside a superconducting solenoid coil that provides
a 1.5 T magnetic field.  Two inner detector configurations were used.  A 2.0 cm radius beampipe with a 3-layer SVD was used for the initial 16\% of the sample and a 1.5 cm radius beampipe with a 4-layer SVD for the rest.  Charged particle identification is accomplished by combining specific ionization measurements in the CDC with the information from an array of aerogel threshold Cherenkov counters, and a barrel-like arrangement of time-of-flight scintillation counters. The analysis procedure is established using Monte Carlo (MC) simulated samples. Particle decays are modeled by the EvtGen package~\cite{evtgen}, with the simulation of detector response performed with GEANT3~\cite{geant3}.

We select charged tracks originating from the collision region with $|dr| < 0.5~\mathrm{cm}$
and $|dz|<2.0~\mathrm{cm}$, where $dr$ and $dz$ are the impact parameters with respect to the nominal interaction point in the plane transverse and parallel to the $e^{+}$ beam, respectively.  We require these charged tracks to have at least two associated hits in the SVD, in both the $z$ and azimuthal projections. Charged hadrons are identified with a likelihood ratio $\textit{L}(K/\pi) =
\textit{L}_{K}/(\textit{L}_{K} + \textit{L}_{\pi})$, where $L_{\pi}$ and $L_{K}$ are the individual likelihood values for the $\pi^{\pm}$ and $K^{\pm}$ hypothesis based on all the available particle identification information. We require $\textit{L}(K/\pi) > 0.6$ and $L(K/\pi) < 0.4$ for $K^{\pm}$ and $\pi^{\pm}$ candidates, respectively. The $K^{0}_S$ candidates are reconstructed from pairs of oppositely charged tracks (assumed to be pions) that form a common vertex, and are identified with an artificial neural network \cite{NN} that combines seven kinematic variables of the $K^{0}_S$ including the finite flight length for $K^{0}_{S}$ vertex from the $e^{+}e^-$ interaction point. More details on $K^{0}_{S}$ identification can be found in Ref. \cite{nisKS}. The invariant mass of the selected candidates is required to satisfy $487~\mathrm{MeV}/c^{2} < M_{K^{0}_S} < 508~\mathrm{MeV}/c^{2}$ that corresponds to approximately three standard deviations ($\sigma$) in mass resolution. The $K^{0}_S$ purity is 96\% after all the $K^{0}_S$ selections are applied. $\pi^{0}$ meson candidates are reconstructed from photon pairs.  Photons are contiguous regions of energy deposit in the ECL without any associated charged tracks. The ratio of the energy deposited in the central $3\times 3$ array of crystals relative to that in the central $5\times 5$ array of crystals is required to be greater than 0.75. The energy of each photon must be greater than 50, 100, and 150 MeV in the barrel region, forward, and backward endcap, respectively. The $\pi^{0}$ momentum is required to be greater than $300~\mathrm{MeV}/c$, and its invariant mass is required to be in the range $120~\mathrm{MeV}/c^{2} < M_{\gamma\gamma} < 148~\mathrm{MeV}/c^{2}$, which corresponds to approximately $\pm3\sigma$ around the nominal $\pi^0$ mass \cite{pdg}.

As the $\omega$ lifetime is negligible, we determine the $D^0$ decay vertex from a kinematic fit constraining the $K^{0}_{S}$, $\pi^+$, $\pi^-$, and $\pi^0$ candidates to come from a common vertex. We constrain the $\pi^0$ mass in this fit by introducing a large uncertainty of 1.0 cm on its vertex position. We select $D^0 \to K^{0}_S \pi^+ \pi^- \pi^0$ candidates in the $\omega$ mass region by requiring  $750 ~\mathrm{MeV}/c^{2} < M_{\pi\pi\pi^0} < 810~\mathrm{MeV}/c^{2}$ that corresponds to approximately
$\pm3 \sigma$ in resolution around the nominal $\omega$ mass \cite{pdg}. The purity of the $\omega$ sample after all selection criteria is 91.4\%. We retain a $D^0 \to K^{0}_S \pi^+ \pi^- \pi^0$ candidate if its invariant mass is in the range $1.80~\mathrm{GeV}/c^{2} < M_{D} < 1.92~\mathrm{GeV}/c^{2}$, and a $D^0 \to K^{-} \pi^+$ candidate if its invariant mass is in the range
$1.83~\mathrm{GeV}/c^{2} < M_{D} < 1.90~\mathrm{GeV}/c^{2}$. The tighter requirement in the latter case is due to better mass resolution. The $D^{*+}$ candidates are reconstructed from the selected $D^0$ and $\pi^{+}_{\rm slow}$ candidates requiring the mass difference between $D^{*+}$ and $D^0$ to lie in the range $m_{\pi^+} < \Delta M < 150~\mathrm{MeV}/c^{2}$. Here, $\pi^{+}_{\rm slow}$ is the charged pion whose momentum tends to be low compared to the final-state particles originating from the $D^0$ decay, and $m_{\pi^+}$ is the charged pion nominal mass~\cite{pdg}. In order to suppress combinatorial background further and veto $D^0$ mesons coming from $B$ decays, the $D^{*+}$ momentum in the center-of-mass frame is required to be greater than $2.55~\mathrm{GeV}/c$. 

The production vertex of the $D^0$, i.e.\ the $D^{*+}$ vertex, is obtained by constraining the $D^0$ momentum to the interaction region (IR). The $\pi^{+}_{\rm slow}$ candidate is refitted to the $D^{*+}$ vertex to improve resolution of $\Delta M$. 
As the IR position varies with changing accelerator conditions, we update the mean position every 10,000 hadronic events. The IR position resolution is determined by comparing the mean IR position with the true production vertex position using MC. The mean width of the IR is $3.34$~mm along the $z$ axis and $82~\mu$m in the horizontal and $4.3~\mu$m in the vertical directions. To further improve vertex resolutions, we require confidence levels to exceed $10^{-3}$ for both fits. After applying all selection criteria there are on average 1.40 (1.01) candidates per event in the $D^{0} \to K^{0}_{S}\omega$ ($K\pi$) decay.
We retain the one having the minimum $\chi^{2}$ value determined from the $\rm \pi_{slow}$ vertex fit. 

The proper decay time of $D^0$ candidates is calculated by projecting the flight length vector connecting the $D^{*+}$ and $D^0$ decay vertices along the direction of the momentum vector $\vec{p}$, and then dividing by the magnitude of $\vec{p}$ and multiplying by the $D^0$ mass. The error on the proper decay time, $\sigma_t$, is calculated from the error matrix of the production vertex position, the decay vertex position, and the momentum $\vec{p}$. The diagonal elements correspond to the variances in these quantities, whereas the off-diagonal elements give the correlations among their uncertainties. The resolution on the decay time is ~$\rm 310~fs$ for $D^{0} \to K^{0}_{S}\omega$ decays, and ~$\rm 162~fs$ 
for $D^{0} \to K\pi$ decays. For both samples, a loose requirement $\sigma_t < 900 ~\rm fs$ is imposed. The worsening in resolution in the $D^{0} \to K^{0}_{S}\omega$ case is due to the presence of $\pi^0$ and $K^{0}_{S}$  in the final state.

According to MC simulation, the selected events can be grouped into the following four categories: signal, random $\pi_{\rm slow}$ background composed of correctly reconstructed $D^0$ mesons combined with a misreconstructed $\pi_{\rm slow}$, combinatorial background, and background due to partially reconstructed multibody charm decays. We first perform a two-dimensional (2D) unbinned maximum-likelihood fit to the variables $(M_{D}, \Delta M)$ in order to extract signal and background fractions. These are then used in the lifetime fits to normalize different lifetime components.

The probability density functions (PDFs) of different event categories are parametrized as follows. For the $D^0 \to K_{S}^{0}\omega$ decay mode, the signal distribution in $M_D$ is modeled  with the sum of
a Crystal Ball $\rm(CB)$ function \cite{CB} and three Gaussian functions all constrained to a common mean, while the distribution in $\Delta M$ is parametrized with the sum of two Gaussian functions constrained to a common mean (double Gaussian function) to describe the core, and the sum of 
an asymmetric Gaussian function and a $\rm CB$ function to model the tails. To account for a correlation between the core widths of $\Delta M$ and $M_{D}$, we  parametrize the former with a second-order polynomial of $|M_D - m_{D^0}|$, where $m_{D^0}$ is the nominal mass~\cite{pdg} of the $D^0$ meson.

The signal distribution of the $D^0 \to K^{-}\pi^{+}$ decay mode is parametrized in $M_{D}$ with a sum of a $\rm CB$ function, a double Gaussian function, and an asymmetric Gaussian function, while in $\Delta M$ it is modeled with a double Gaussian function to describe the core, and with a sum of a $\rm CB$ function and two asymmetric Gaussian functions to describe the tails. The correlation between the core widths of $\Delta M$ and $M_{D}$ is parametrized as for the $D^0 \to K_{S}^{0}\omega$ mode.

The distribution of random $\rm \pi_{slow}$ background is peaking in $M_D$ and smooth in
$\Delta M$. The former is parametrized with the signal PDF and the latter with a threshold function
\begin{equation}
  F_{\textrm{thr}}(Q) = Q^\alpha e^{-\beta Q}, ~~~ Q > 0,
  \label{thr_func}
\end{equation}
where $Q \equiv \Delta M - m_{\pi^+}$, and $\alpha$ and $\beta$ are two shape parameters. 

The distribution of combinatorial background is smooth in both variables. We parametrize
it in $M_D$ with either a first-order polynomial ($K^{-}\pi^+$) or a second-order polynomial ($K_{S}^{0}\omega$); and in $\Delta M$ with the threshold function as in Eq.~(\ref{thr_func}). 

The background due to partially reconstructed multibody charm decays is smooth in $M_D$ but exhibits a broad peak in $\Delta M$. In the case of $K_{S}^{0}\omega$, this background is small (about 3\% of the total background) and its shape in $M_D$ is very similar to that of the combinatorial background. We decide to combine this background with the combinatorial background by adding an additional Gaussian term to the parametrization in $\Delta M$. The parameters of this additional function and its fraction are fixed from the fit to MC simulation. In the case of $K^{-}\pi^+$, we treat this background separately.
The distribution is parametrized with an exponential function in $M_D$, and with a double Gaussian function in $\Delta M$ whose parameters are fixed to values obtained from MC simulation.

The robustness of our fitting model is tested with MC samples that corresponds to the Belle data set in integrated luminosity. The obtained signal and background fractions in the signal region, defined in Table~\ref{tab:regions}, are consistent with the ones determined with MC ``truth matching;'' the difference between the two is, in all cases, within one standard deviation.

\begin{table}[htbp]
  \begin{center}
    \caption{Definitions of signal region and sidebands.
      Units are $\mathrm{GeV}/c^2$.} 
    \begin{tabular}{c|c}
      \hline
      \hline
      \multicolumn{2}{c}{Signal region} \\
      \hline
      $K_{S}^{0}\omega$ & $K^{-}\pi^{+}$\\
      $1.84 < M_{D} < 1.885$ & $1.85 < M_{D} < 1.88$ \\  \hline
      \multicolumn{2}{c}{$0.144 < \Delta M < 0.147$} \\      \hline\hline
      \multicolumn{2}{c}{Sidebands} \\
      \hline
      $K_{S}^{0}\omega$ & $K^{-}\pi^{+}$\\
      $1.76< M_{D} < 1.79$ & $1.76 < M_{D} < 1.80$ \\
      $1.92< M_{D} < 1.95$ & $1.91 < M_{D} < 1.95$ \\ \hline
      \multicolumn{2}{c}{$m_{\pi^+} < \Delta M < 0.142$} \\
      \multicolumn{2}{c}{$0.149 < \Delta M < 0.150$} \\      
      \hline
      \hline
    \end{tabular}
    \label{tab:regions}
  \end{center}
\end{table}

\begin{figure}[htbp]
  \begin{center}
    \includegraphics[width=0.5\textwidth]{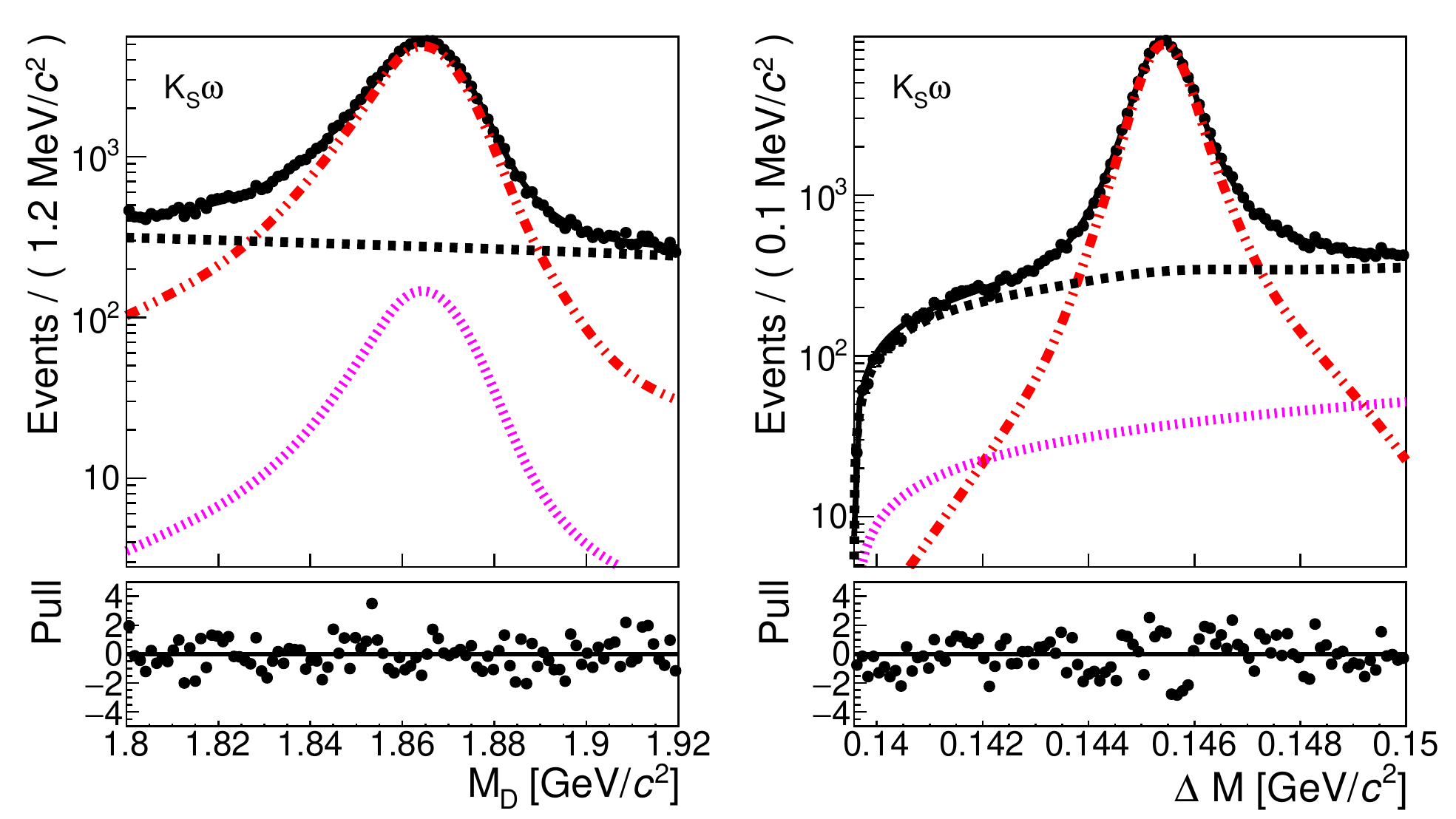}
    \includegraphics[width=0.5\textwidth]{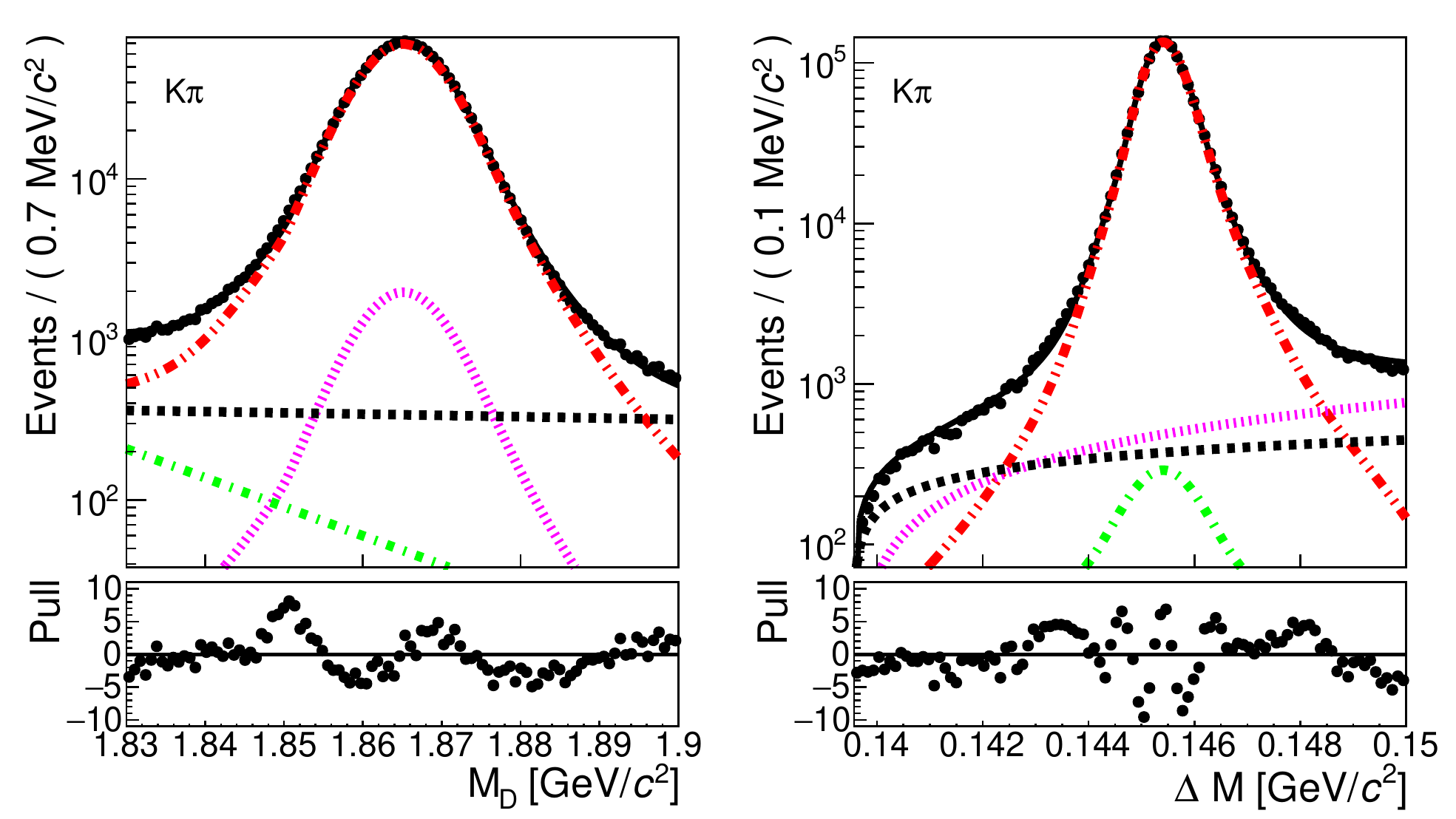}
    \caption{ [color online]. Projections of the 2D fit on $M_D$ (left) and $\Delta M$
      (right) for $D^0 \to K_{S}^{0}\omega$ (top) and  $D^0 \to K^{-}\pi^{+}$ (bottom).
      Points with error bars represent the data. The curves show projections of fitted
      PDF: total PDF projection in solid black,
      signal contribution in double-dot-dashed red,
      combinatorial background in dashed black,
      random $\pi_{\rm slow}$ background in dotted magenta,
      and multibody backgound as dash-dotted green. (The total PDF is hard to see as it closely follows the data points.)}
    \label{fig:projections}
  \end{center}
\end{figure}

After validating the fitting model, we proceed to fit the data sample. The results are shown in Fig.~\ref{fig:projections} and are listed in Table~\ref{tab:datayields}. We measure the signal fractions of 96.3\% ($K_{S}^{0}\omega$) and 99.6\% ($K^{-}\pi^{+}$) by integrating events in the signal region. 

\begin{table}[htbp]
  \begin{center}
    \caption{Yields from the 2D fit to data.} 
    \begin{tabular}{lcc}\hline\hline
      $K_{S}^{0}\omega$ components & Full region   & Signal region \\
      \hline
      Signal                       & 107978$\pm$455   & 90930 \\
      Random $\pi_{\rm slow}$ background           &   3238$\pm$346   &   918 \\
      Combinatorial background      &  27793$\pm$447   &  3554 \\
      \hline
      \hline
      $K^{-}\pi^{+}$ components    & Full region   & Signal region \\
      \hline
      Signal                       & 1507830$\pm$1310 & 1375245 \\
      Random $\pi_{\rm slow}$ background           &   42899$\pm$459  &   13380 \\
      Combinatorial background      &   33828$\pm$384  &    4620 \\
      Multibody background             &    6769$\pm$415  &    1686 \\
      \hline
      \hline
    \end{tabular}
    \label{tab:datayields}
  \end{center}
\end{table}

Finally we perform unbinned maximum-likelihood fits for lifetime using the events in the signal region. We parametrize the proper decay-time distribution as
\begin{equation}
  F(t; \tau) = \frac{f_{\rm sig}}{\tau} \int e^{-t'/\tau} R(t - t') dt' + (1 - f_{\rm sig})B(t),
\end{equation}
where the first term represents signal and the second term background, $f_{\rm sig}$ is the fraction of signal events determined with the 2D fit described earlier, $\tau$ is the effective signal lifetime, and $R(t-t')$ is the resolution function. The resolution function is parametrized with the sum of three ($K_{S}^{0}\omega$) or four ($K^{-}\pi^{+}$) Gaussian functions constrained to the common mean. Besides the effective lifetime $\tau$, the free parameters of the fit are the resolution function mean, the widths, and the fraction of each Gaussian function.

The background term $B(t)$ is parametrized with two lifetime components: a zero-lifetime component corresponding to combinatorial background, and a component with an effective lifetime $\tau_b$ corresponding to multibody charm background:
\begin{equation}
  B(t) = \int [f_0 \delta(t') + \frac{1-f_0}{\tau_b} e^{-t'/\tau_b}] R_{b}(t-t') dt',
\end{equation}
where $f_0$ is the fraction of zero-lifetime component and $R_{b}(t-t')$ is the resolution function for background, parametrized with a sum of three Gaussian functions constrained to the common mean. The parameters of $B(t)$ are obtained by fitting the proper-time distribution of events in the sidebands as defined in Table~\ref{tab:regions}. The sidebands are chosen such that they contain negligible amounts of signal.

The lifetime fitting model is tested with four statistically independent MC samples each corresponding to the integrated luminosity in data. The resulting fitted lifetimes are found to be consistent with the generated value, and $y_{CP}$ determined from the fitted lifetimes of $D^0 \to K_{S}^{0}\omega$ and $D^0 \to K^{-}\pi^{+}$ is compatible with zero within one standard deviation.

\begin{figure}[htbp]
  \begin{center}
    \includegraphics[width=0.4\textwidth]{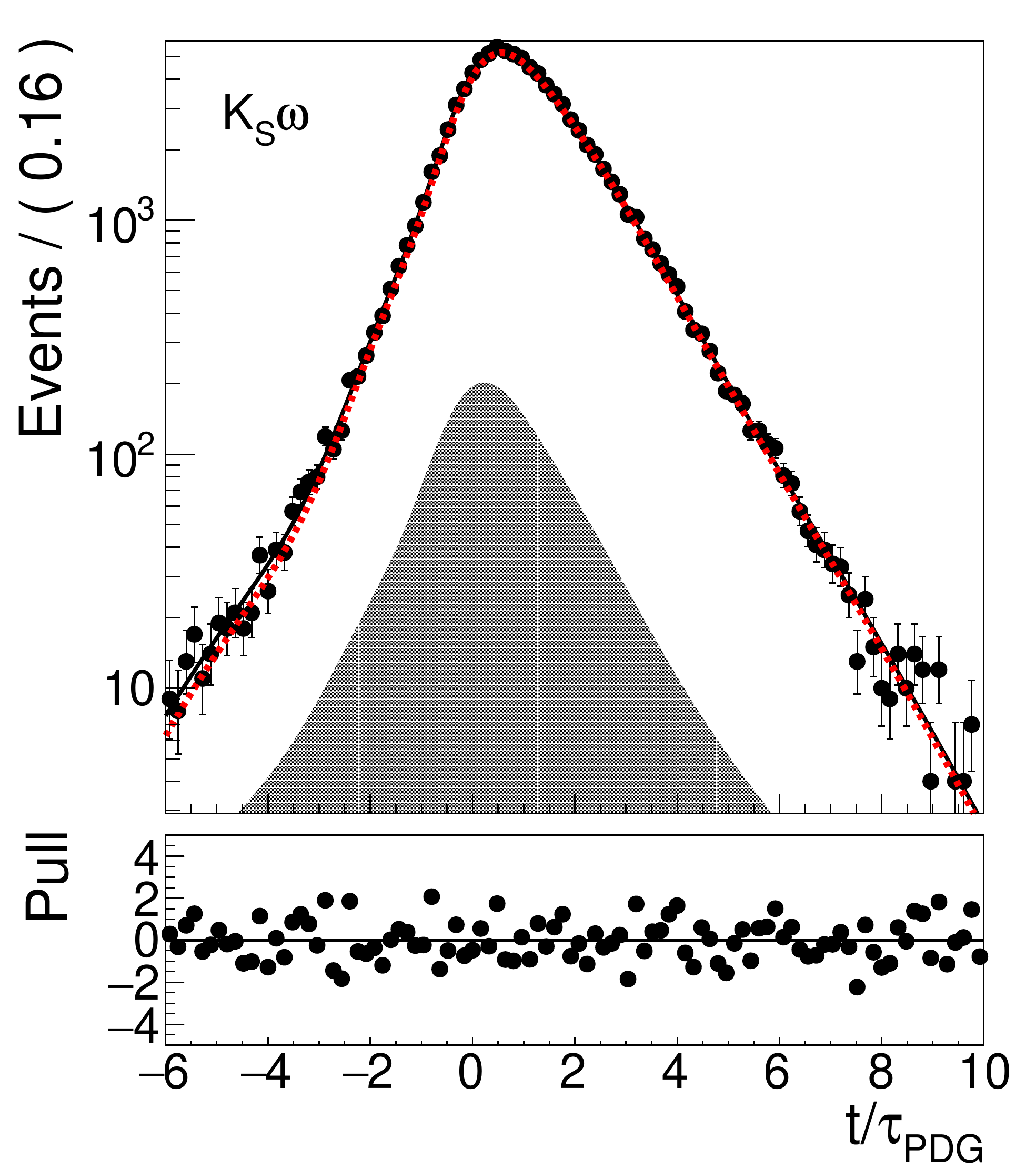}\\
    \includegraphics[width=0.4\textwidth]{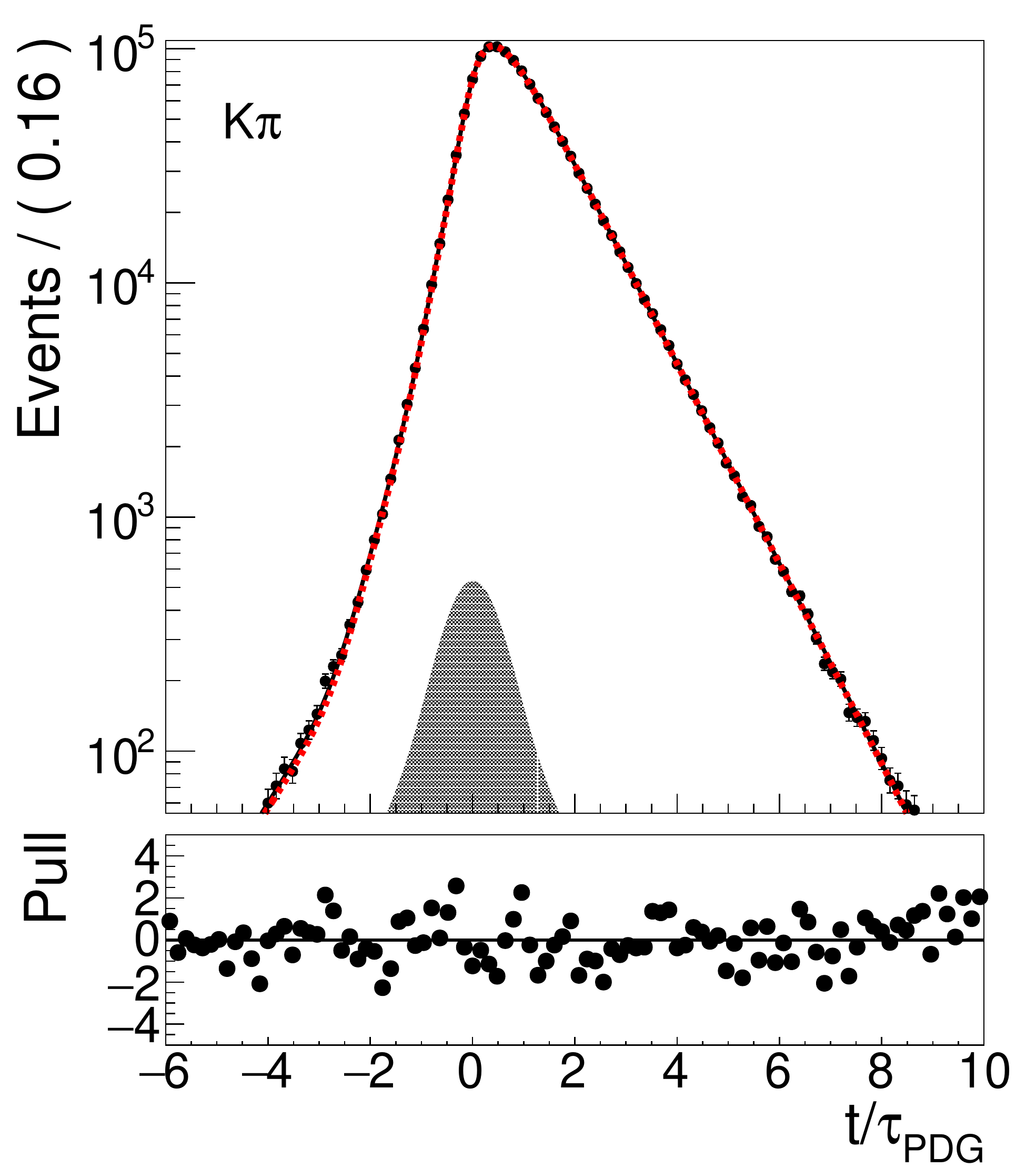}\\
    \caption{ [color online]. Results of the fit to the measured proper decay time
      distributions: (top) $D^0 \to K^{0}_{S}\omega$ and (bottom) $D^0 \to K\pi$.
      Points with error bars represent the data, the solid black curves are the fitted function, the dashed red curves are the signal contribution, and
      the shaded surfaces beneath are the background estimated from sidebands.}
    \label{fig:lifetime_data}
  \end{center}
\end{figure}

Lifetime fits on the data are shown in Fig.~\ref{fig:lifetime_data}. The $\chi^2$ per number of degrees of freedom of the $D^0 \to K_{S}^{0}\omega$ and $D^0 \to K^{-}\pi^{+}$ lifetime fits are 0.90 and 1.10, respectively. We measure $\tau_{K^{0}_S \omega} = (410.47 \pm 3.73) ~\rm fs$ and $\tau_{K\pi} = (406.53 \pm 0.57)~\rm fs$, and $y_{\rm CP}$ = $(0.96\pm0.91)\%$, where the uncertainties are statistical.

Besides $D^{0}\to K^{0}_{S}\omega$ decay, the reconstructed final state $K^{0}_{S} \pi^+ \pi^- \pi^0$ might include contributions from other intermediate resonances, or no resonance at all. Depending on orbital angular momenta, some of these decay modes might be $CP$-even. The presence of $CP$-even component in the signal reduces the measured $y_{CP}$ by a factor of $1 - 2f_{CP+}$, where $f_{CP+}$ is the fraction of $CP$-even decays in the signal component. Since this fraction is not well known in the selected mass region of $\omega$, we assign a systematic uncertainty to the measured $y_{CP}$ by conservatively assuming that all non-$\omega$ decays are $CP$-even. The fraction of non-$\omega$ decays is determined from a fit to the $M_{\pi\pi\pi^0}$ distribution in which the $M_{\pi\pi\pi^0}$ requirement is loosened but events are still required to be in the signal region. The fraction of events under the $\omega$ peak obtained from the fit and corrected for a small amount of random combinations of $\omega$ and $K_S^0$ (2.5\%) is 88.0\%, while the signal fraction from the 2D fit is 96.3\%. From the ratio of the two (91.4\%) we find the upper limit $f_{CP+}=8.6\%$. The systematic uncertainty in $y_{CP}$ due to possible presence of $CP$-even decays in the sample is therefore at most $2f_{CP+}\cdot y_{CP} = +0.17\%$.

Other sources of systematic uncertainties are listed in Table \ref{tab:syst1}. We vary the requirement on the $K^{0}_{\rm S}$ flight length in steps of 0.1 mm up to 1.0 mm; we find no significant bias in the $D^0$ lifetime and assign the maximum variation observed of 0.01\% as the systematic uncertainty in $y_{CP}$. To assign systematics due to different energy thresholds used for different barrel regions, we divide the whole barrel region into three equal bins and assign a maximum energy threshold of each photon of 70 MeV to each bin. We observe an average bias of 0.1\% which we assign as the systematic due to $\pi^0$ reconstruction. We vary our selection criteria on $\sigma_t$ by $\pm 50~{\rm fs}$ and find a 0.21\% variation in $y_{CP}$. Variation of $D$ mass window position and size by $\pm2.5~\mathrm{MeV}/c^2$ leads to a 0.13\% change in $y_{CP}$. 
We vary the signal fraction by its statistical and systematic uncertainties; we find a 0.14\% variation due to statistics and, from MC simulation, 0.10\% due to the fixed shape parameters in the $(M_{D}, \Delta M)$ fit. 
These two contributions are combined in quadrature and the result is assigned as the systematic uncertainty due to the signal fraction.
Note that difference between the data and fit visible in Fig.~\ref{fig:projections} for the $D^0\to K\pi$ mode has a negligible effect on the extracted lifetime.

By choosing different sidebands to obtain the decay-time dependence of background $B(t)$, we find a variation of 0.32\% in $y_{CP}$. We also vary the background lifetime by the lifetime difference obtained in simulation between background events in the signal region and those in the sidebands; we find a variation of 0.03\% in $y_{CP}$. We vary each fixed background shape parameter by its uncertainty; by taking into account correlations among the parameters, we obtain a variation of 0.43\% in $y_{CP}$. By summing the above contributions in quadrature we obtain a total systematic uncertainty of 0.62\%; the systematic uncertainty due to possible presence of $CP$-even decays in the data sample (discussed earlier) is treated separately.
\begin{table}[htbp]
  \begin{center}
    \caption{Summary of absolute systematic uncertainties.}
    \label{tab:syst1}
    \begin{tabular}{lccc}
      \hline
      \hline
      Source & $y_{CP}$ uncertainty [\%]  \\
      \hline
      $K^{0}_{\rm S}$ selection  & $\pm0.01$ \\
      $\pi^0$ reconstruction  & $\pm0.10$ \\
      $\sigma_t$ selection & $\pm0.21$ \\ 
      $M_D$ signal window & $\pm0.13$ \\ 
      Signal fraction   & $\pm0.17$ \\ 
      Sideband selection   & $\pm0.32$ \\ 
      Signal/sideband background differences  & $\pm0.03$ \\ 
      Sideband parametrization & $\pm0.43$ \\ 
      \hline
      Quadrature Sum   & $\pm0.62$ \\ 
      \hline
      $CP$-even decays  & ${}^{+0.17}_{-0.00}$ \\ 
      \hline
      \hline
    \end{tabular}
  \end{center}
\end{table}

In summary, we have measured for the first time the mixing parameter $y_{CP}$ in the $CP$-odd decay $D^0 \to K_S^0 \omega$. We obtain
\begin{equation}
  y_{CP} = (0.96 \pm 0.91 \pm 0.62 {}^{+0.17}_{-0.00})\%,
\end{equation}  
where the first uncertainty is statistical, the second is systematic due to event selection and background, and the last is due to the possible presence of $CP$-even decays in the final state. The result is consistent with our previous measurement in the $CP$-odd decay $D^0 \to K_S^0 \phi$~\cite{zupanc}, as well as with measurements in the $CP$-even decays $D^0 \to K^+K^-$ and $D^0 \to \pi^+\pi^-$~\cite{staric_2016, lees, lhcb_ycp}. The result also agrees with the world average of $y_{CP}$~\cite{hflav}.
In the future, comparing more precise measurements of $y_{CP}$ with that of $y$ may reveal new physics effects in the charm system.

We thank the KEKB group for the excellent operation of the
accelerator; the KEK cryogenics group for the efficient
operation of the solenoid; and the KEK computer group, and the Pacific Northwest National
Laboratory (PNNL) Environmental Molecular Sciences Laboratory (EMSL)
computing group for strong computing support; and the National
Institute of Informatics, and Science Information NETwork 5 (SINET5) for
valuable network support.  We acknowledge support from
the Ministry of Education, Culture, Sports, Science, and
Technology (MEXT) of Japan, the Japan Society for the 
Promotion of Science (JSPS), and the Tau-Lepton Physics 
Research Center of Nagoya University; 
the Australian Research Council including grants
DP180102629, 
DP170102389, 
DP170102204, 
DP150103061, 
FT130100303; 
Austrian Science Fund (FWF);
the National Natural Science Foundation of China under Contracts
No.~11435013,  
No.~11475187,  
No.~11521505,  
No.~11575017,  
No.~11675166,  
No.~11705209;  
Key Research Program of Frontier Sciences, Chinese Academy of Sciences (CAS), Grant No.~QYZDJ-SSW-SLH011; 
the  CAS Center for Excellence in Particle Physics (CCEPP); 
the Shanghai Pujiang Program under Grant No.~18PJ1401000;  
the Ministry of Education, Youth and Sports of the Czech
Republic under Contract No.~LTT17020;
the Carl Zeiss Foundation, the Deutsche Forschungsgemeinschaft, the
Excellence Cluster Universe, and the VolkswagenStiftung;
the Department of Science and Technology of India; 
the Istituto Nazionale di Fisica Nucleare of Italy; 
National Research Foundation (NRF) of Korea Grant
Nos.~2016R1\-D1A1B\-01010135, 2016R1\-D1A1B\-02012900, 2018R1\-A2B\-3003643,
2018R1\-A6A1A\-06024970, 2018R1\-D1A1B\-07047294, 2019K1\-A3A7A\-09033840,
2019R1\-I1A3A\-01058933;
Radiation Science Research Institute, Foreign Large-size Research Facility Application Supporting project, the Global Science Experimental Data Hub Center of the Korea Institute of Science and Technology Information and KREONET/GLORIAD;
the Polish Ministry of Science and Higher Education and 
the National Science Center;
the Ministry of Science and Higher Education of the Russian Federation, Agreement 14.W03.31.0026; 
the Slovenian Research Agency;
Ikerbasque, Basque Foundation for Science, Spain;
the Swiss National Science Foundation; 
the Ministry of Education and the Ministry of Science and Technology of Taiwan;
and the United States Department of Energy and the National Science Foundation.



\begin{thebibliography}{99}
\bibitem{conjugate} Charge-conjugate modes are implicitly included unless stated otherwise.
\end{thebibliography}
\end{document}